\documentclass[twocolumn,preprintnumbers,amsmath,amssymbm,prd]{revtex4}
\usepackage{epsfig}
\usepackage{graphicx}

\begin{document}
\title{Bulk emission by higher-dimensional black holes: almost perfect blackbody
radiation}
\author{Shahar Hod}
\affiliation{The Ruppin Academic Center, Emeq Hefer 40250, Israel}
\affiliation{ } \affiliation{The Hadassah Institute, Jerusalem
91010, Israel}
\date{\today}

\begin{abstract}
\ \ \ We study the Hawking radiation emitted into the bulk by
$(D+1)$-dimensional Schwarzschild black holes. It is well-known that
the black-hole spectrum departs from exact blackbody form due to the
frequency dependence of the `greybody' factors. For intermediate
values of $D$ ($3\leq D\lesssim10$), these frequency-dependent
factors may significantly modify the spectrum of the emitted
radiation. However, we point out that for $D\gg1$, the typical
wavelengths in the black-hole spectrum are much {\it shorter} than
the size of the black hole. In this regime, the greybody factors are
well described by the geometric-optics approximation according to
which they are almost frequency-independent. Following this
observation, we argue that for higher-dimensional black holes with
$D\gg1$, the total power emitted into the bulk should be well
approximated by the analytical formula for perfect blackbody
radiation. We test the validity of this analytical prediction with
numerical computations.
\end{abstract}
\bigskip
\maketitle


\section{Introduction}
Models with large extra dimensions are widely regarded as the most
promising candidates for a consistent unified theory of the
fundamental forces \cite{Ark,Ran}. In these models it is usually
assumed that the standard model fields are confined to a
four-dimensional hypersurface (known as the `brane'), but gravity
(and possibly scalar fields) is free to propagate in a
higher-dimensional compact space (the `bulk').

Such higher-dimensional models are intriguing because they suggest
an elegant resolution for the so-called hierarchy problem. In
particular, they may explain why gravity is perceived to be much
weaker than the other forces. According to these higher-dimensional
models, the traditional Planck scale, $M_P$, is only an effective
energy scale derived from the fundamental higher-dimensional one,
$M_*$, through the relation \cite{Ark,Ran,Kanti1,KanDo}
\begin{equation}\label{Eq1}
M^2_P\sim M^{2+n}_*R^n\  ,
\end{equation}
where $R$ and $n$ are the size and number of extra dimensions,
respectively. From (\ref{Eq1}) one deduces that if the volume of the
compact space, $V\sim R^n$, is large (i.e if $R\gg \ell_P$, where
$\ell_P\approx 10^{-35}m$ is the traditional Planck-length), then
the $(4+n)$-dimensional Planck mass, $M_*$, will be much lower than
the $4$-dimensional one, $M_P$. Remarkably, by lowering the Planck
scale $M_*$ closer to the energy scale of modern accelerators, the
possibility of producing miniature black holes during high-energy
scattering processes now becomes more realistic \cite{Kanti1,Gid}.

If the horizon of the formed black hole is much smaller than the
size of the extra dimensions, $r_H\ll R$, then the produced black
hole may be considered as a higher-dimensional object that is
submerged into the extra-dimensional spacetime \cite{Kanti1}. If
created, these mini black holes are expected to evaporate quickly by
the emission of thermal Hawking radiation \cite{Haw}. It is hoped
that this characteristic radiation could be detected in future
high-energy experiments. If detected, this radiation may provide an
experimental verification of the celebrated Hawking evaporation
process.

A higher-dimensional black hole emits radiation both in the bulk and
on the brane. It is usually assumed that only gravitons (and
possibly scalar fields) can propagate in the bulk. Thus, these are
the only types of fields allowed to be emitted in the bulk during
the Hawking evaporation phase \cite{Kanti1}. It is important to
realize that for an observer located on the brane, the radiation
emitted in the bulk will be perceived as a missing energy signal. On
the other hand, radiation on the brane may be detected directly.
Nevertheless, in order to have a complete picture of the
characteristics of the radiation spectrum on the brane, it is
important to know how much energy is emitted (lost) in the bulk
\cite{Kanti1}.

The non-trivial spacetime exterior to the black-hole horizon is
characterized by an effective scattering potential. This potential
barrier scatters part of the outgoing radiation back into the black
hole \cite{Kanti1}. As a consequence of this radiation
backscattering, the power spectrum that would be detected by an
observer at spatial infinity would not be universal. In particular,
it would depend on several parameters: the energy $\omega$ of the
emitted particle, its spin $s$, and the dimensionality $(D+1)$ of
spacetime \cite{Kanti1} (We denote by $D=3+n$ the total number of
spatial dimensions). The dependence of the emission spectrum on all
these parameters is encoded into the `greybody factor'
$\sigma(\omega)_{sD}$. This factor acts as a filtering function
which characterizes the interaction of the emitted quanta with the
curvature scattering potential which surrounds the black hole. This
interaction modifies the thermal radiation spectrum \cite{Kanti1}
[see Eq. (\ref{Eq7}) below.]

The greybody factors can be calculated analytically in the
low-energy $\omega r_H\ll1$ and high-energy $\omega r_H\gg1$ regimes
{\cite{Kanti1,CarCavHal}. However, for moderate values of $D$ most
of the Hawking radiation is actually emitted around $\omega
r_H\approx 1$, where the analytical approximations breakdown. Thus,
{\it numerical} integration of the perturbed field equations seems
necessary in order to compute the exact greybody factors and to find
the corresponding black-hole emission power
\cite{Kanti1,Kono,KanDo,Page,HarKan}.

Nevertheless, in this paper we point out that for higher-dimensional
spacetimes with $D\gg1$, the typical wavelengths emitted into the
bulk are much {\it shorter} than the size of the black hole. In this
regime, the greybody factors are well described by the
geometric-optics approximation. As a consequence, we shall show
below that for higher-dimensional black holes with $D\gg1$, the
total power emitted into the bulk is well approximated by the {\it
analytical} formula for perfect (undistorted) blackbody radiation.

\section{Hawking radiation in the bulk}
We consider higher-dimensional black holes that have horizon radius
much smaller than the size of the extra dimensions, $r_H\ll R$.
These mini black holes are completely submerged into a
$(D+1)$-dimensional spacetime that, to a very good approximation,
has one timelike and $D$ non-compact spacelike coordinates
\cite{Kanti1}. If we further assume that the black hole is
spherically-symmetric with ADM mass ${\cal M}$, the spacetime
outside the horizon is described by the $(D+1)$-dimensional
Schwarzschild-Tangherlini metric \cite{SchTang,Kun} (we use natural
units in which $G=c=1$):
\begin{equation}\label{Eq2}
ds^2=-H(r)dt^2+{H(r)}^{-1}dr^2+r^2d\Omega^{(D-1)}\ ,
\end{equation}
where
\begin{equation}\label{Eq3}
H(r)=1-{\Big({r_H\over r}\Big)}^{D-2}\  .
\end{equation}
Here
\begin{equation}\label{Eq4}
r_H={\Big[{{16\pi {\cal M}}\over{(D-1)A_{D-1}}}\Big]}^{1\over{D-2}}
\end{equation}
is the black hole's radius and
\begin{equation}\label{Eq5}
A_{D-1}={{2\pi^{D/2}}\over {\Gamma(D/2)}}
\end{equation}
is the area of a unit $(D-1)$-sphere (The black hole's area is given
by $A_H=A_{D-1}r_H^{D-1}$.) The Hawking temperature of the black
hole is given by \cite{Kanti1}
\begin{equation}\label{Eq6}
T={{(D-2)\hbar}\over{4\pi r_H}}\  .
\end{equation}

For one helicity degree of freedom, the energy emitted per unit time
into the bulk by a $(D+1)$-dimensional black hole is given by
\cite{Kanti1}:
\begin{equation}\label{Eq7}
P_D=\sum_{j}{\int_0^{\infty}}\sigma(\omega)_{sjD} {{\hbar\omega\
dV_D(\omega)}\over{{(e^{\hbar\omega/T}-1)}{(2\pi)^D}}}\  ,
\end{equation}
where $s$ is the spin of the emitted quantum ($s=2$ for gravitons
and $s=0$ for scalars), $j$ its angular momentum quantum number,
$\sigma(\omega)_{sjD}$ is the frequency-dependent greybody factor of
the spacetime, and
\begin{equation}\label{Eq8}
dV_D(\omega)=[2\pi^{D/2}/\Gamma(D/2)]\omega^{D-1}d\omega\
\end{equation}
is the volume in frequency space of the shell
$(\omega,\omega+d\omega)$. Substituting (\ref{Eq8}) into
(\ref{Eq7}), one obtains
\begin{equation}\label{Eq9}
P_D={{2\pi^{D/2}}\over{(2\pi)^D\Gamma(D/2)}}\sum_{j}{\int_0^{\infty}}\sigma(\omega)_{sjD}
{{\hbar\omega^D\ d\omega}\over{{(e^{\hbar\omega/T}-1)}}}\
\end{equation}
for the power radiated into the bulk per one degree of freedom by
the $(D+1)$-dimensional black hole.

As discussed above, the factor $\sigma(\omega)_{sjD}$ is not
universal -- it has a complicated dependence on several parameters
of the system: the energy $\omega$ of the emitted particle, its spin
$s$, its angular momentum $j$, and the number $D$ of spatial
dimensions. Thus, this factor modifies the radiation spectrum that
reaches an observer at spatial infinity. In particular, such an
observer would not detect a perfect thermal radiation. The curvature
potential barrier which surrounds the black hole mostly blocks the
low energy part ($\omega r_H\ll1$) of the emission spectrum (this
should be contrasted with pure black body radiation in a flat
spacetime). As a consequence, the black-hole power spectrum is
expected to peak at higher frequencies as compared to those of
perfect blackbody radiation with the same temperature.

We point out that the distribution
$\omega^{D}/(e^{\hbar\omega/T}-1)$ in Eq. (\ref{Eq9}) peaks at the
characteristic frequency
\begin{equation}\label{Eq10}
\omega^*={{DT}\over{\hbar}}[1-e^{-D}+O(e^{-2D})]\  .
\end{equation}
Taking cognizance of Eq. (\ref{Eq6}) for the black hole's
temperature, one finds for $D\gg1$
\begin{equation}\label{Eq11}
\omega^* r_H={{D(D-2)}\over{4\pi}}\gg1\  .
\end{equation}
This implies that for higher-dimensional black holes with $D\gg1$,
the typical wavelengths in the Hawking radiation are much {\it
shorter} than the size of the black hole. Since for $D\gg1$ the
integral in (\ref{Eq9}) is dominated by large frequencies around
$\omega^*$, one may approximate $P_{D}$ by
\begin{equation}\label{Eq12}
P_D\simeq{{2\pi^{D/2}}\over{(2\pi)^D\Gamma(D/2)}}\sum_{j}\sigma(\omega^*)_{sjD}{\int_0^{\infty}}
{{\hbar\omega^D\ d\omega}\over{{(e^{\hbar\omega/T}-1)}}}\  .
\end{equation}

In the short wavelength regime $\omega r_H\gg1$ (the geometrical
optics limit), geometric arguments \cite{Kanti1,Misner,Myers,San}
show that the absorption cross-section
$\sigma_{\text{abs}}\equiv\sum_{j}\sigma(\omega)_{sjD}$ is a
constant {\it independent} of $\omega$ and $s$ \cite{Kanti1}:
Consider a massless particle in a circular orbit around a black hole
described by the line-element (\ref{Eq2}). Its equation of motion
$p^{\mu}p_{\mu}=0$ takes the form \cite{Kanti1}
\begin{equation}\label{Eq13}
\Big({1\over r}
{{dr}\over{d\phi}}\Big)^2={{1}\over{b^2}}-{{H(r)}\over{r^2}}\  ,
\end{equation}
where $b$ is the ratio of the angular momentum of the particle over
its linear momentum. Since the left-hand-side of (\ref{Eq13}) is
positive definite, the classically accessible regime of the particle
is defined by the relation $b<\text{min}(r/\sqrt{H})$. Thus, the
closest distance the particle can get from the black hole is given
by \cite{Kanti1,Myers}
\begin{equation}\label{Eq14}
b=r_c\equiv
\Big({{D}\over{2}}\Big)^{{1}\over{D-2}}\sqrt{{{D}\over{D-2}}}r_H\  .
\end{equation}

The radius $r_c$ defines the absorptive area of the black hole at
high energies. For large values of the energy of the scattered
particle, the greybody factor $\sigma_{abs}$ becomes equal to the
area of an absorptive body of radius $r_c$ which is projected on a
plane parallel to the orbit of the moving particle
\cite{Kanti1,Misner}:
\begin{equation}\label{Eq15}
\sigma_{\text{abs}}={{2\pi}\over{D-1}}{{\pi^{{D-3}\over{2}}}\over{\Gamma({{D-1}\over2})}}r_c^{D-1}\
.
\end{equation}
Note that $\sigma_{\text{abs}}$ can also be written as
\begin{equation}\label{Eq16}
\sigma_{\text{abs}}={{1}\over{\sqrt{\pi}(D-1)}}{{\Gamma({D\over2})}\over{\Gamma({{D-1}\over2})}}
\Big({{r_c}\over{r_H}}\Big)^{D-1}A_H\  .
\end{equation}
In \cite{Kanti1,Corn} it was demonstrated by explicit numerical
computations that both the total absorption cross-section of
gravitational perturbations (composed of tensor, vector, and scalar
type perturbations \cite{CarCavHal}) and the absorption
cross-section of scalar fields tend to the classical expression
(\ref{Eq16}) in the high-energy regime $\omega r_H\gg1$.

Substituting (\ref{Eq6}) and (\ref{Eq16}) into (\ref{Eq12}), and
using the relation
\begin{equation}\label{Eq17}
{\int_0^{\infty}}{{x^Ddx}\over{e^x-1}}=\zeta(D+1)\Gamma(D+1)\  ,
\end{equation}
where $\zeta(z)$ is the Riemann zeta function, one finds
\begin{equation}\label{Eq18}
P_D^{\text{tot}}\simeq
N_D\Big({{D-2}\over{4\pi}}\Big)^{D+1}\Big({{r_c}\over{r_H}}\Big)^{D-1}{{D\zeta(D+1)\hbar}\over{\pi
r_H^2}}\
\end{equation}
for the total power radiated into the bulk from a
$(D+1)$-dimensional black hole with $D\gg1$, where $N_D$ is the
effective number of massless degrees of freedom (the number of
polarization states). Massless scalars contribute $1$ to $N_D$,
while gravitational waves contribute $(D+1)(D-2)/2$ to $N_D$
\cite{CarCavHal}.

Note that for $D\gg1$ one has $(r_c/r_H)^{D-1}\to De/2$. Thus, the
total radiated power (\ref{Eq18}) can be approximated by the compact
formula
\begin{equation}\label{Eq19}
P_D^{\text{tot}}\simeq{{8\pi N_D\hbar}\over{e
r_H^2}}\Big({{D}\over{4\pi}}\Big)^{D+3}\  .
\end{equation}

\section{Analytical vs. numerical results}
It is of interest to test the validity of the approximated
analytical formula (\ref{Eq18}) for the energy emission rate into
the bulk from a $(D+1)$-dimensional Schwarzschild black hole. Our
analytical treatment is based on the observation (\ref{Eq11})
according to which the typical wavelengths in the Hawking radiation
are much {\it shorter} than the size of the black hole in the
$D\gg1$ regime: $\omega^* r_H={{D(D-2)}\over{4\pi}}\gg1$. In this
regime, the geometric-optics approximation predicts a frequency-{\it
independent} greybody factors given by Eq. (\ref{Eq16}). This
implies that the dominant part of the black-hole emission spectrum
(around $\omega^*$) is hardly affected by the curvature potential
barrier in the $D\gg1$ regime. For this reason, we expect the
black-hole emission properties (in the $D\gg1$ regime) to be well
approximated by the emission properties of a perfect blackbody with
the same temperature.

We shall first compare the location of the peak in the black-hole
emission spectrum with the value predicted by the blackbody
analytical expression (\ref{Eq10}). In Table \ref{Table1} we display
the ratio
$\Omega_D\equiv{{\omega^*_{\text{blackhole}}}\over{\omega^*_{\text{blackbody}}}}$
between the numerically computed
\cite{Kanti1,Kono,KanDo,Page,HarKan} peak-frequency and the
analytical prediction (\ref{Eq10}). We present results for the total
gravitational spectrum and for the scalar spectrum in the bulk. As
explained above, the curvature potential barrier outside the black
hole mostly affect the low-frequency part of the emission spectrum.
The result is that the black-hole power spectrum is expected to peak
at higher frequencies as compared to those of perfect blackbody
radiation with the same temperature. From Table \ref{Table1} one
indeed finds
$\omega^*_{\text{blackhole}}>\omega^*_{\text{blackbody}}$.

For scalar waves the agreement
$\omega^*_{\text{blackhole}}\simeq\omega^*_{\text{blackbody}}$
($\Omega_D\simeq1$) is quite impressive already at $D=3$. This
indicates that the dominant part of the scalar emission spectrum
(around $\omega^*$) is hardly affected by the curvature potential
barrier for all $D$ values. We therefore expect the scalar emission
power to follow closely the blackbody analytical expression
(\ref{Eq18}) for all $D$ values. Below we shall confirm this
expectation.

For gravitational waves one finds a large deviation between the peak
values $\omega^*_{\text{blackhole}}$ and
$\omega^*_{\text{blackbody}}$ in the case of three spatial
dimensions. We therefore expect the gravitational emission rate to
be suppressed as compared to the blackbody analytical prediction
(\ref{Eq18}) for $D=3$. Below we shall confirm this expectation.
However, for $D=10$ one finds a good agreement between
$\omega^*_{\text{blackhole}}$ and $\omega^*_{\text{blackbody}}$.
This indicates that for $D\gg1$, the dominant part of the
gravitational emission spectrum is hardly affected by the curvature
potential barrier.

\begin{table}[htbp]
\centering
\begin{tabular}{|c|c|c|}
\hline $D$ & $\Omega_D(\text{gravitational})$ &
$\Omega_D(\text{scalar})$ \\
\hline
\ $3$\ \ &\ $1.590$\ \ & $1.058$ \\
\ $10$\ \ &\ $1.028$\ \ & $1.017$ \\
\hline
\end{tabular}
\caption{The ratio
$\Omega_D\equiv\omega^*_{\text{blackhole}}/\omega^*_{\text{blackbody}}$
between the numerically computed peak-frequency and the approximated
analytical prediction (\ref{Eq10}). We present results for the total
gravitational spectrum and for the scalar spectrum in the bulk. Due
to the presence of the curvature potential barrier surrounding the
black hole, one expects to find $\Omega_D>1$ with $\Omega_D\to1$ for
$D\gg1$.} \label{Table1}
\end{table}

We shall now compare the total power emitted into the bulk from a
$(D+1)$-dimensional black hole with the power predicted by the
blackbody analytical model. In Table \ref{Table2} we display the
ratio $\Pi_D\equiv
{{P_{\text{blackhole}}}\over{P_{\text{blackbody}}}}$ between the
numerically computed \cite{Kanti1,Kono,KanDo,Page,HarKan} black-hole
emission power and the blackbody analytical expression (\ref{Eq18}).
We present results for the total emission of gravitational waves and
scalar waves in the bulk.

For scalar waves, the agreement between the black-hole numerical
results and the blackbody analytical prediction is quite impressive
already at $D=3$. This confirms our earlier expectation. For
gravitational waves, the black-hole emission power for $D=3$ is
suppressed as compared to the blackbody analytical prediction
(\ref{Eq18}). Again, this confirms our earlier expectation. However,
one learns from Table \ref{Table2} that the agreement between the
black-hole numerical data and the blackbody analytical formula
(\ref{Eq18}) improves considerably as the number $D$ of spatial
dimensions increases. This is to be expected, since the larger is
the number of spatial dimensions, the larger are the characteristic
frequencies which dominate the emission spectrum, see Eq.
(\ref{Eq11}). Thus, the larger is the number of spatial dimensions,
the smaller is the relative part of the emission spectrum which is
blocked by the curvature potential barrier which surrounds the black
hole.

\begin{table}[htbp]
\centering
\begin{tabular}{|c|c|c|}
\hline $D$ & $\Pi_D(\text{gravitational})$ &
$\Pi_D(\text{scalar})$ \\
\hline
\ $3$\ \ &\ $0.027$\ \ & $1.037$ \\
\ $4$\ \ &\ $0.247$\ \ & $0.974$ \\
\ $5$\ \ &\ $0.543$\ \ & $0.945$ \\
\ $6$\ \ &\ $0.715$\ \ & $0.934$ \\
\ $7$\ \ &\ $0.783$\ \ & $0.931$ \\
\ $8$\ \ &\ $0.814$\ \ & $0.929$ \\
\ $9$\ \ &\ $0.833$\ \ & $0.922$ \\
\ $10$\ \ &\ $0.846$\ \ & $0.928$ \\
\hline
\end{tabular}
\caption{The ratio $\Pi_D\equiv
P_{\text{blackhole}}/P_{\text{blackbody}}$ between the numerically
computed black-hole emission power and the blackbody analytical
expression (\ref{Eq18}). We present results for the total emission
of gravitational waves and scalar waves in the bulk. One expects to
find $\Pi_D\to 1$ for $D\gg1$.} \label{Table2}
\end{table}

\section{Summary}
We have studied the Hawking radiation emitted into the bulk by
$(D+1)$-dimensional Schwarzschild black holes. It is well-known that
the black-hole spectrum departs from exact blackbody form. The
emission spectrum is influenced by the frequency dependence of the
greybody factors [see Eq. (\ref{Eq9})]. For the canonical case of
three spatial dimensions, the frequency-dependent greybody factors
must be computed {\it numerically} in order to obtain the exact
black-hole emission power \cite{Page,San}. This is also the
situation for intermediate values of $D$ \cite{Kanti1,Kono,KanDo}.

However, in this paper we have pointed out that for $D\gg1$, the
typical wavelengths in the bulk spectrum are much {\it shorter} than
the size of the black hole. In this regime, the greybody factors are
well described by the geometric-optics approximation. According to
this approximation, the greybody factors are frequency-independent;
they are simply given by the projected area of an absorptive body of
radius $r_c$. This implies that for higher-dimensional Schwarzschild
black holes with $D\gg1$, the total power emitted into the bulk is
well described by the analytical formula (\ref{Eq18}) of perfect
blackbody radiation. We have tested this prediction and found a
reasonably good agreement already at $D\simeq10$ between the
(numerically computed) black-hole power and the (analytically
calculated) blackbody power.

As emphasized in \cite{CarCavHal}, the relative contributions of the
higher partial waves to the emission power increase with $D$. Thus,
contributions from high values of $l$ are needed in order to obtain
accurate {\it numerical} results for large values of $D$. For
example, in four dimensions the contribution of the $l=2$ mode is
two orders of magnitude larger than the contribution of the $l=3$
mode. However, in ten dimensions the first $10$ modes must be
considered for a meaningful numerical result (see \cite{CarCavHal}).
Therefore, precise numerical values for very large number of spatial
dimensions require the most CPU-time. This fact limits the utility
of the numerical computations to moderate values of $D$. In fact,
the numerical results that appear in the literature are limited to
the case $D\leq10$. Luckily, for higher values of $D$ the agreement
between the analytical approximation (\ref{Eq18}) and the exact
results is expected to be very good. Thus, an analytical formula
like (\ref{Eq18}) allows the calculation of the emission power in
cases which would otherwise require long numerical integration
times.

\bigskip
\noindent
{\bf ACKNOWLEDGMENTS}

This research is supported by the Meltzer Science Foundation. I
thank Yael Oren, Arbel M. Ongo, and Liran Shimshi for helpful
discussions.


\end{document}